# Refining Focus in AI for Lung Cancer: Comparing Lesion-Centric and Chest-Region Models with Performance Insights from Internal and External Validation

**Fakrul Islam Tushar**

Dept. of Electrical & Computer Engineering, Pratt School of Engineering, Duke University, Durham.
Center for Virtual Imaging Trials, Carl E. Ravin Advanced Imaging Laboratories, Department of Radiology, Duke University School of Medicine, Durham, NC.
**Contact:** tushar.ece@duke.edu

## Abstract

**Background:** AI-based classification models are essential for improving lung cancer diagnosis. However, the relative performance of lesion-level versus chest-region models in internal and external datasets remains unclear.

**Purpose:** This study evaluates the performance of lesion-level and chest-region models for lung cancer classification, comparing their effectiveness across internal Duke Lung Nodule Dataset 2024 (DLND24) and external (LUNA16, NLST) datasets, with a focus on subgroup analyses by demographics, histology, and imaging characteristics.

**Materials and Methods:** Two AI models were trained: one using lesion-centric patches ($64\times64\times64$; $x, y, z$) and the other using chest-region patches ($512\times512\times8$; $x, y, z$). Internal validation was conducted on DLND24, while external validation utilized LUNA16 and NLST datasets. The models' performances were assessed using AUC-ROC, with subgroup analyses for demographic, clinical, and imaging factors. Statistical comparisons were performed using DeLong's test. Gradient-based visualizations and probability distribution were further used for analysis.

**Results:** The lesion-level model consistently outperformed the chest-region model across datasets. In internal validation, the lesion-level model achieved an AUC of 0.71 (95% CI: 0.61–0.81), compared to 0.68 (95% CI: 0.57–0.77) for the chest-region model. External validation showed similar trends, with AUCs of 0.90 (95% CI: 0.87–0.92) and 0.81 (95% CI: 0.79–0.82) on LUNA16 and NLST, respectively. Subgroup analyses revealed significant advantages for lesion-level models in certain histological subtypes (adenocarcinoma) and imaging conditions (CT manufacturers).

**Conclusion:** Lesion-level models demonstrate superior classification performance, especially for external datasets and challenging subgroups, suggesting their clinical utility for precision lung cancer diagnostics.

## Introduction

Lung cancer remains the leading cause of cancer-related deaths globally, with early detection being paramount to improving survival rates [1-4]. Low-dose computed tomography (LDCT) is a standard modality for lung cancer screening, offering high sensitivity for detecting pulmonary nodules [4]. However, the radiological interpretation of CT scans is time-intensive and prone to variability[4]. Artificial intelligence (AI) has emerged as a promising tool to assist in lung cancer diagnosis by automating the detection and classification of nodules [5-9].

While lesion-level AI models analyze specific nodule-centric regions, chest-region models encompass a broader anatomical context, potentially incorporating additional diagnostic information [7, 9, 10]. However, a comprehensive comparison of these approaches across diverse datasets and patient subgroups is lacking. This study aims to evaluate the performance of lesion-level and chest-region models for lung cancer classification using internal and external datasets, with subgroup analyses to explore variations by demographics, histology, and imaging parameters. Grad-CAM and related visualization techniques analyze feature attribution.

## Materials and Methods

### Patient Data and Imaging Datasets
This study utilized three datasets: an internal dataset, Duke Lung Nodule Dataset 2024 (DLND24), and two external datasets, LUNA16 and NLST used in an earlier study [7, 11]. DLND24 consists of annotated chest CT scans from a single institution with lesion-level 3D bounding boxes. LUNA16 includes 888 CT scans annotated for pulmonary nodules, while NLST comprises low-dose CT scans from a large-scale cancer screening trial [7, 8, 12]. Table 1 shows the Demographic distribution of the data Cohort used for development (DLND24) and test (DLND24, LUAN16, NLST) sets.

### Study Design and Workflow
Two AI models were developed and tested: one using lesion-level patches and another using chest-region inputs. Models were trained on DLND24 and validated on both internal and external datasets (LUNA16, NLST). Subgroup analyses assessed model performance by demographics, clinical factors, and imaging characteristics.

### Image Preprocessing
For the lesion-level model, patches (64×64×64; $x, y, z$) were extracted from CT scans, centered on nodules. For the chest-region model, 3D volumes (512×512×8; $x, y, z$) spanning the entire lung region were used. All CT volumes were resampled to a resolution of $0.7 \times 0.7 \times 1.25$ ($x, y, z$). Intensity values were clipped between -1000 and 500 and standardized to a mean of 0 and a standard deviation of 1.

### AI Model Development
Two AI models were developed for lung cancer classification. The lesion-level model employed the ResNet50-SWS++, an architecture introduced in an earlier study, which utilized Strategic WarmStart++ (SWS++) pretraining [7]. This model was trained end-to-end using nodule-centric patches (64×64×64; $x, y, z$) for malignancy classification. The chest-region model leveraged the pre-trained weights of a false positive reduction model, as detailed in prior work [7]. These weights served as initialization for the chest-region lung cancer classification model, which analyzed chest-centric patches (512×512×8; $x, y, z$). Both models were optimized using binary cross-entropy loss and trained with the Adam optimizer.

### Grad-CAM and Feature Localization
Grad-CAM and Grad-CAM++ were applied to generate heatmaps highlighting areas contributing to malignancy predictions [13]. Additional techniques, including SmoothGrad and Guided Backpropagation (GuidedBG), were used to validate and refine feature attribution [14]. Visualizations were qualitatively assessed to identify differences in feature localization between the two models.

### Probability Distribution
Distributions were analyzed to evaluate model confidence and decision boundaries for in-distribution (ID) and external (out-of-distribution; OOD) cases. KDE plots were generated for probability. The degree of overlap between distributions served as an indicator of the model's discriminative power and confidence in predictions.

### Evaluation Metrics
Model performance was assessed using the Area under the curve-curve-receiver operating Characteristic (AUC-ROC) [15]. Subgroup analyses stratified results by gender, race, histology, smoking status, and CT manufacturer. Confidence intervals were derived using bootstrapping. DeLong's test was used for statistical comparisons between models [15].

## Results

### Overall Performance
The lesion-level model consistently outperformed the chest-region model across all datasets. In internal validation using the DLND24 dataset, the lesion-level model achieved an AUC of 0.71 (95% CI: 0.61–0.81), while the chest-region model achieved an AUC of 0.68 (95% CI: 0.57–0.77) (Fig.2a). External validation further highlighted the strength of the lesion-level model in Fig.2b and Fig.2c. On the LUNA16 dataset, the lesion-level model demonstrated exceptional performance with an AUC of 0.90 (95% CI: 0.87–0.92), significantly surpassing the chest-region model, which had an AUC of 0.63 (95% CI: 0.58–0.67). Similarly, on the NLST dataset, the lesion-level model achieved an AUC of 0.81 (95% CI: 0.79–0.82), compared to 0.71 (95% CI: 0.69–0.72) for the chest-region model. These results indicate that the lesion-level approach is more effective at capturing malignancy-specific features.

### Subgroup Analysis
The subgroup analyses revealed key insights into model performance across different patient demographics, histological subtypes, and imaging parameters shown in **Fig.3**. For demographic groups, the lesion-level model showed higher AUCs for both male and female patients compared to the chest-region model. Among smoking subgroups, the lesion-level model demonstrated significant advantages for current smokers, with an AUC improvement of approximately 10%. Histological analysis indicated that the lesion-level model performed exceptionally well for adenocarcinoma, with an AUC of 0.85, and squamous cell carcinoma, with an AUC of 0.81, compared to consistently lower performance by the chest-region model across all subtypes. Imaging parameter analysis on the LUNA16 dataset highlighted the robustness of lesion-level models across different CT manufacturers, including GE and Siemens, further validating their generalizability (**Fig.4**).

### Probability Distribution
Probability distributions showed (**Fig.5**) sharper separations between benign and malignant cases for the lesion-level model compared to the chest-region model. The lesion-level model displayed well-separated peaks for labels 0 (no cancer) and 1 (cancer), indicating high confidence and discriminative power. In contrast, the chest-region model exhibited significant overlap between benign and malignant distributions, suggesting lower confidence and reduced ability to differentiate between classes effectively.

### Grad-CAM Visualization
Grad-CAM analysis revealed distinct differences in feature localization between the two models (**fig.6**). The lesion-level model consistently focused on nodules, with heatmaps from Grad-CAM and Grad-CAM++ showing strong activations in nodule-specific regions. SmoothGrad and GuidedBG further validated these findings, demonstrating consistent and localized feature attribution. In contrast, the chest-region model displayed broader and less focused activations. Heatmaps from Grad-CAM and Grad-CAM++ showed diffuse activation patterns, often highlighting regions outside of nodules. This lack of specificity was corroborated by SmoothGrad and GuidedBG visualizations, which revealed inconsistent feature attribution.

## Discussion

The results demonstrate the clear advantage of lesion-level models over chest-region models in lung cancer classification. The superior performance of lesion-level models in both internal and external validation suggests that focusing on nodule-specific features enables these models to better differentiate between benign and malignant lesions. The enhanced AUCs in external datasets, particularly LUNA16 and NLST, highlight the generalizability of the lesion-level approach. Additionally, subgroup analyses showed that lesion-level models consistently outperformed chest-region models across diverse demographics, histological subtypes, and imaging parameters, further underscoring their robustness.

These findings are also consistent with the weakly-supervised classification studies associated with the classification of present or absent nodules [16-19].

The superior performance of lesion-level models has significant clinical implications. By leveraging nodule-specific information, these models align more closely with radiologists' diagnostic workflows, where the primary focus is on evaluating nodule characteristics such as size, shape, and margins. The lesion-level models' ability to generalize across datasets, including external validation cohorts, suggests they are well-suited for real-world applications. Furthermore, their higher sensitivity in certain subgroups indicates their potential to address variability in patient populations and disease presentations. Despite the promising results, several limitations warrant discussion. The datasets used in this study may introduce selection bias, as they include only annotated CT scans with labeled nodules, which may not represent the full spectrum of cases encountered in clinical practice. Additionally, while qualitative Grad-CAM and KDE analyses provide insights, quantitative metrics for interpretability and decision boundary robustness are needed. Building on these findings, future work will focus on integrating multimodal data, such as radiomics biomarkers and clinical history, to provide a more comprehensive assessment of malignancy risk. Exploring the inclusion of longitudinal imaging data could also enhance the ability to track disease progression and provide dynamic risk predictions.

Lesion-level models demonstrate superior performance and feature localization compared to chest-region models for lung cancer classification. Visualizations highlight their focus on clinically relevant nodule features, reinforcing their suitability for nodule-specific diagnostics. These findings emphasize the importance of lesion-centric approaches in AI-driven lung cancer screening.

## Acknowledgment

This work was funded in part by the Center for Virtual Imaging Trials, NIH/NIBIB P41-EB028744, and Putnam Vision Award awarded by Duke Radiology. I also suggest acknowledging the Duke Lung Cancer Screening Program.

## Dataset and Code Availability

The dataset used in this study is the publicly available Duke Lung Cancer Screening Dataset (DLCSD) at Zenodo: https://zenodo.org/records/10782891. The code for data preprocessing, segmentation, feature extraction, model training, and evaluation will be openly available at GitHub: https://github.com/fitushar/AI-in-Lung-Health-Benchmarking-Detection-and-Diagnostic-Models-Across-Multiple-CT-Scan-Datasets

**Table 1. Demographic distribution of the data Cohort used for training, development and test sets.**

| Category | | All (%) | Training (%) | Validation (%) | Testing (%) |
|---|---|---|---|---|---|
| | | | | | |
| **Duke Lung Cancer Screening Dataset** | | | | | |
| **Gender** | | | | | |
| | Male | 811 (50.28) | 559 (52.48) | 167 (46.78) | 85 (42.93) |
| | Female | 802 (49.72) | 499 (47.16) | 190 (53.22) | 113 (57.07) |
| | | | | | |
| **Age** | Mean (min-max) | 66 (50-89) | 66 (50-89) | 66 (55-78) | 66 (54-79) |
| | | | | | |
| **Race** | White | 1,195 (74.09) | 775 (73.25) | 280 (78.43) | 140 (70.71) |
| | Black/AA | 366 (22.69) | 247 (23.35) | 68 (19.05) | 51 (25.76) |
| | Other/Unknown | 52 (3.22) | 36 (3.40) | 9 (2.52) | 7 (3.54) |
| | | | | | |
| **Ethnicity** | | | | | |
| | Not Hispanic | 1,555 (96.40) | 1,019 (96.31) | 344 (96.36) | 192 (96.97) |
| | Unavailable | 52 (3.22) | 35 (3.31) | 12 (3.36) | 5 (2.53) |
| | Hispanic | 6 (0.37) | 4 (0.38) | 1 (0.28) | 1 (0.51) |
| | | | | | |
| **Smoking status** | | | | | |
| | Current | 826 (53.92) | 538 (53.48) | 189 (56.08) | 99 (52.38) |
| | Former | 704 (45.95) | 467 (46.42) | 147 (43.62) | 90 (47.62) |
| | Other/Unknown | 2 (0.13) | 1 (0.10) | 1 (0.30) | |
| | | | | | |
| **Cancer** | | | | | |
| | Patient | | | | |
| | Benign | 1,469 (91.07) | 965 (91.21) | 324 (90.76) | 180 (90.91) |
| | Malignant | 144 (8.93%) | 93 (8.79) | 33 (9.24) | 18 (9.09) |
| | | | | | |
| | Lung-RADS | | | | |
| | 1 | 8 (0.64) | 5 (0.61) | 2 (0.73) | 1 (0.64) |
| | 2 | 703 (56.20) | 463 (56.33) | 152 (55.68) | 88 (56.41) |
| | 3 | 219 (17.51) | 143 (17.40) | 49 (17.95) | 27 (17.31) |
| | 4A | 165 (13.19) | 106 (12.90) | 38 (13.92) | 21 (13.46) |
| | 4B | 113 (9.03) | 78 (9.49) | 21 (7.69) | 14 (8.97) |
| | 4X | 43 (3.44) | 27 (3.28) | 11 (4.03) | 5 (3.21) |
|  | Nodule |  |  |  |  |
|  | Benign | 2,223 (89.38) | 1,452 (89.74) | 510 (88.70) | 261 (88.78) |
|  | Malignant | 264 (10.62) | 166 (10.26) | 65 (11.30) | 33 (11.22) |
|  |  |  |  |  |  |
|  | Lung-RADS |  |  |  |  |
|  | 1 | 10 (0.52) | 5 (0.61) | 2 (0.73) | 1 (0.64) |
|  | 2 | 970 (50.18) | 463 (56.33) | 152 (55.68) | 88 (56.41) |
|  | 3 | 374 (19.35) | 143 (17.40) | 49 (17.95) | 27 (17.31) |
|  | 4A | 278 (14.38) | 106 (12.90) | 38 (13.92) | 21 (13.46) |
|  | 4B | 216 (11.17) | 78 (9.49) | 21 (7.69) | 14 (8.97) |
|  | 4X | 85 (4.40) | 27 (3.28) | 11 (4.03) | 5 (3.21) |
|  |  |  |  |  |  |
|  | **National Lung Screening Trial (NLST)** |  |  |  |  |
| Gender |  |  |  |  |  |
|  | Male | 572 (59.03) |  |  | 572 (59.03) |
|  | Female | 397 (40.97) |  |  | 397 (40.97) |
|  |  |  |  |  |  |
| Age | Mean (min-max) | 63 (55-74) |  |  | 63 (55-74) |
|  |  |  |  |  |  |
| Race | White | 900 (92.88) |  |  | 900 (92.88) |
|  | Black/AA | 43 (4.44) |  |  | 43 (4.44) |
|  | Other/Unknown | 26 (2.68) |  |  | 26 (2.68) |
|  |  |  |  |  |  |
| Ethnicity |  |  |  |  |  |
|  | Not Hispanic | 954 (98.45) |  |  | 954 (98.45) |
|  | Unavailable | 7 (0.72) |  |  | 7 (0.72) |
|  | Hispanic | 8 (0.83) |  |  | 8 (0.83) |
|  |  |  |  |  |  |
| Smoking status |  |  |  |  |  |
|  | Current | 535 (55.21) |  |  | 535 (55.21) |
|  | Former | 434 (44.79) |  |  | 434 (44.79) |
|  |  |  |  |  |  |
| Pack-year smoking history |  |  |  |  |  |
|  | 21-30 years | 18 (1.86) |  |  | 18 (1.86) |
|  | > 30+ years | 951 (98.14) |  |  | 951 (98.14) |

| Study year of the last screening | | | | | |
|---|---|---|---|---|---|
| | Year 0 | 265 (27.35) | | | 265 (27.35) |
| | Year 1 | 282 (29.10) | | | 282 (29.10) |
| | Year 2 | 422 (43.55) | | | 422 (43.55) |
| **Cancer** | | | | | |
| | Patient | | | | |
| | Malignant (Screen-detected) | 926 (95.56) | | | 926 (95.56) |
| | Malignant (Other) | 43 (4.44) | | | 43 (4.44) |
| | Nodule | | | | |
| | Malignant (Screen-detected) | 1,143 (95.89) | | | 1,143 (95.89) |
| | Malignant (Other) | 49 (4.11) | | | 49 (4.11) |
| | | LUNA16 | | | |
| Gender | N/A | | | | N/A |
| Age | N/A | | | | N/A |
| Nodule Annotations | Patients | 601 (100) | | | 601 (100) |
| | Nodule | 1186 (100) | | | 1186 (100) |
| Radiologist-Visual Assessed Malignancy Index' (RVAMI) | | | | | |
| | Nodule | | | | |
| | Positive | 327 (48.3) | | | 327 (48.3) |
| | Negative | 350 (51.7) | | | 350 (51.7) |

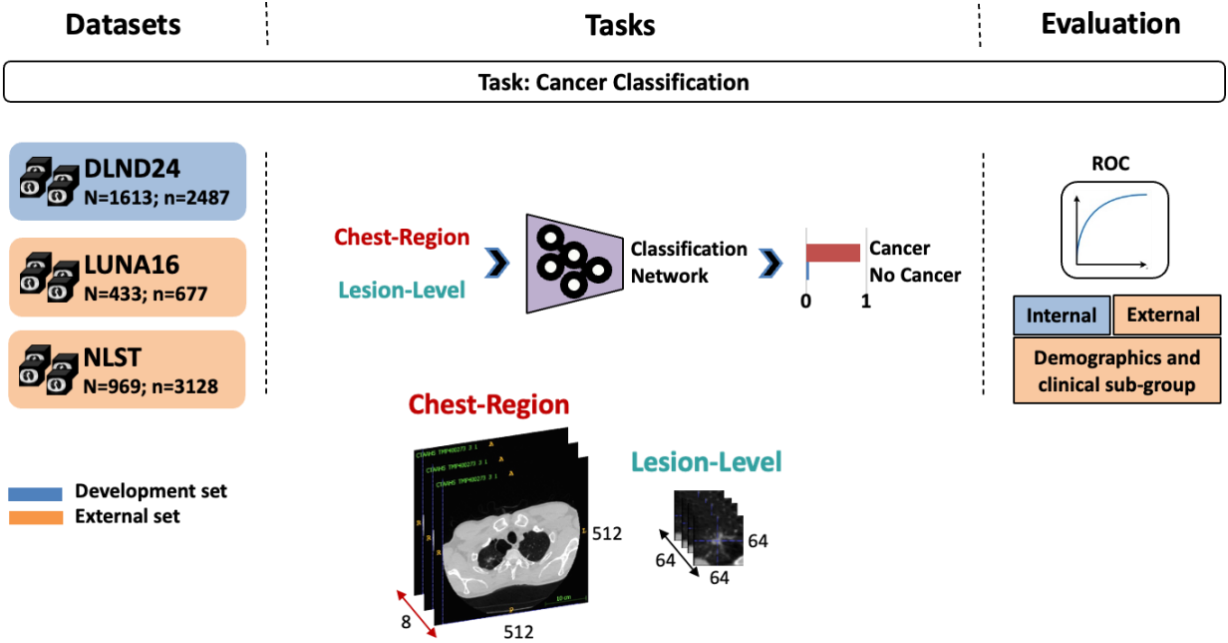

**Figure 1.** Overview of the study design and methodology for evaluating lesion-level and chest-region models for lung cancer classification. The analysis utilized three datasets: DLND24 (internal dataset, N=1613; n=2487 nodules), LUNA16 (external dataset, N=433; n=677 nodules), and NLST (external dataset, N=969; n=3128 nodules). Both models were trained on DLND24 and evaluated on internal and external test sets. The **chest-region model** analyzed large chest patches (512×512×8 voxels), while the **lesion-level model** focused on nodule-centric patches (64×64×64 voxels). The models were assessed for classification performance using AUC-ROC curves and subgroup analyses based on demographic and clinical factors. The workflow highlights the comparative evaluation of both models for internal, external, and subgroup-based performance assessments, emphasizing interpretability and clinical relevance.

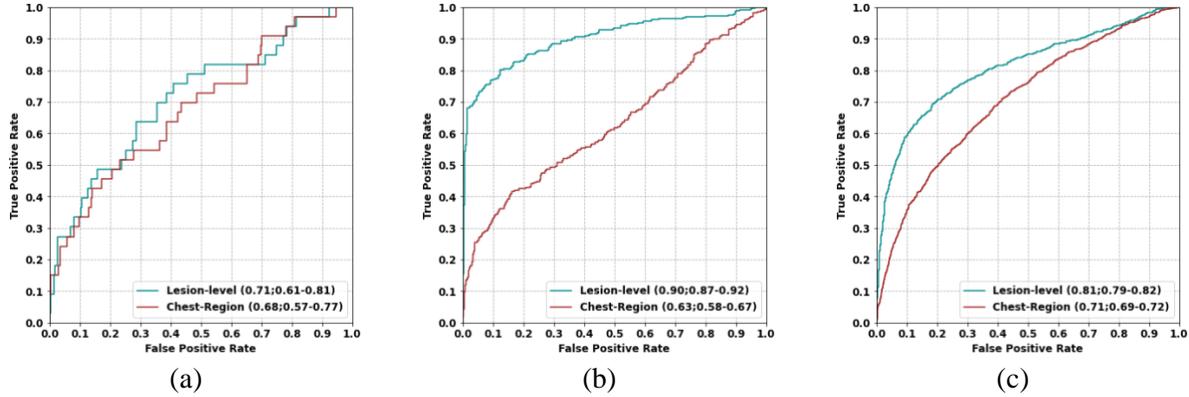

**Figure 2.** Performance of Lesion-Level and Chest-Region Models Across Datasets. (a) Internal dataset performance evaluated on DLND24, showing AUC-ROC comparisons for lesion-level and chest-region models. (b) External dataset performance on LUNA16, highlighting differences in model classification accuracy. (c) External dataset performance on NLST, illustrating generalizability of both models across a large screening dataset.

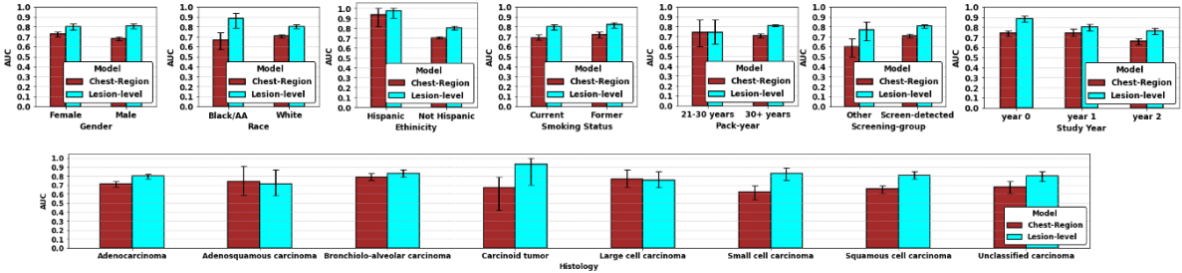

**Figure 3.** Subgroup performance comparison of lesion-level and chest-region models on the NLST test set. The bar plots display the AUC-ROC performance of both models across various demographic, clinical, and histological subgroups, with error bars representing the 95% confidence intervals. The top row illustrates performance differences based on gender (female, male), race (Black/AA, White), ethnicity (Hispanic, Non-Hispanic), and smoking status (current smoker, former smoker). The middle row stratifies results by pack-years (21–30 years, 30+ years), screening group (screen-detected, other), and study year (year 0, year 1, year 2). The bottom row focuses on histological subtypes. Lesion-level models consistently outperform chest-region models across most subgroups, particularly in challenging histological cases such as adenocarcinoma and small cell carcinoma, underscoring their robustness and clinical utility.

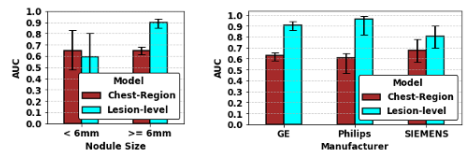

**Figure 4.** Comparative AUC-ROC performance of lesion-level and chest-region models on the LUNA16 dataset, stratified by nodule size. The bar plots represent model performance for small nodules (<6 mm) and large nodules (≥6 mm), with error bars indicating the 95% confidence intervals. The lesion-level model demonstrates consistently higher performance than the chest-region model across both size categories, with a significant advantage observed for large nodules (≥6 mm). These findings highlight the lesion-level model's superior ability to capture malignancy-relevant features across varying nodule sizes within the LUNA16 dataset.

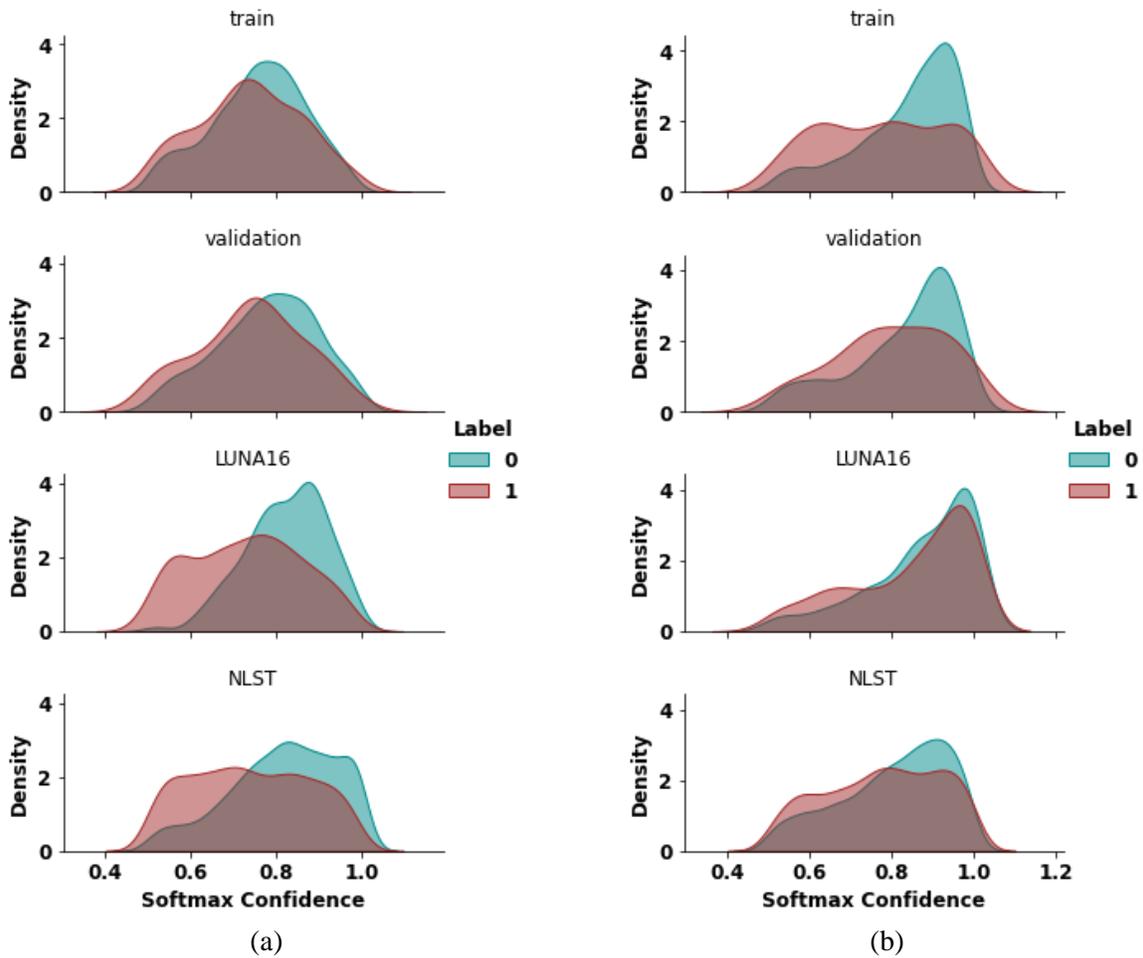

**Figure 5.** Probability distribution of cancer (label 1) and no cancer (label 0) predictions for lesion-level and chest-region models. Kernel density estimation (KDE) plots show the predicted probabilities for the LUNA16 test set and NLST test set. The lesion-level model demonstrates sharp and well-separated probability distributions for cancer and no cancer predictions across both datasets, indicating high discriminative power and confidence in classification. In contrast, the chest-region model shows overlapping probability distributions, reflecting reduced confidence and a diminished ability to differentiate between cancerous and non-cancerous cases. These findings highlight the lesion-level model's robustness and reliability in producing accurate predictions.

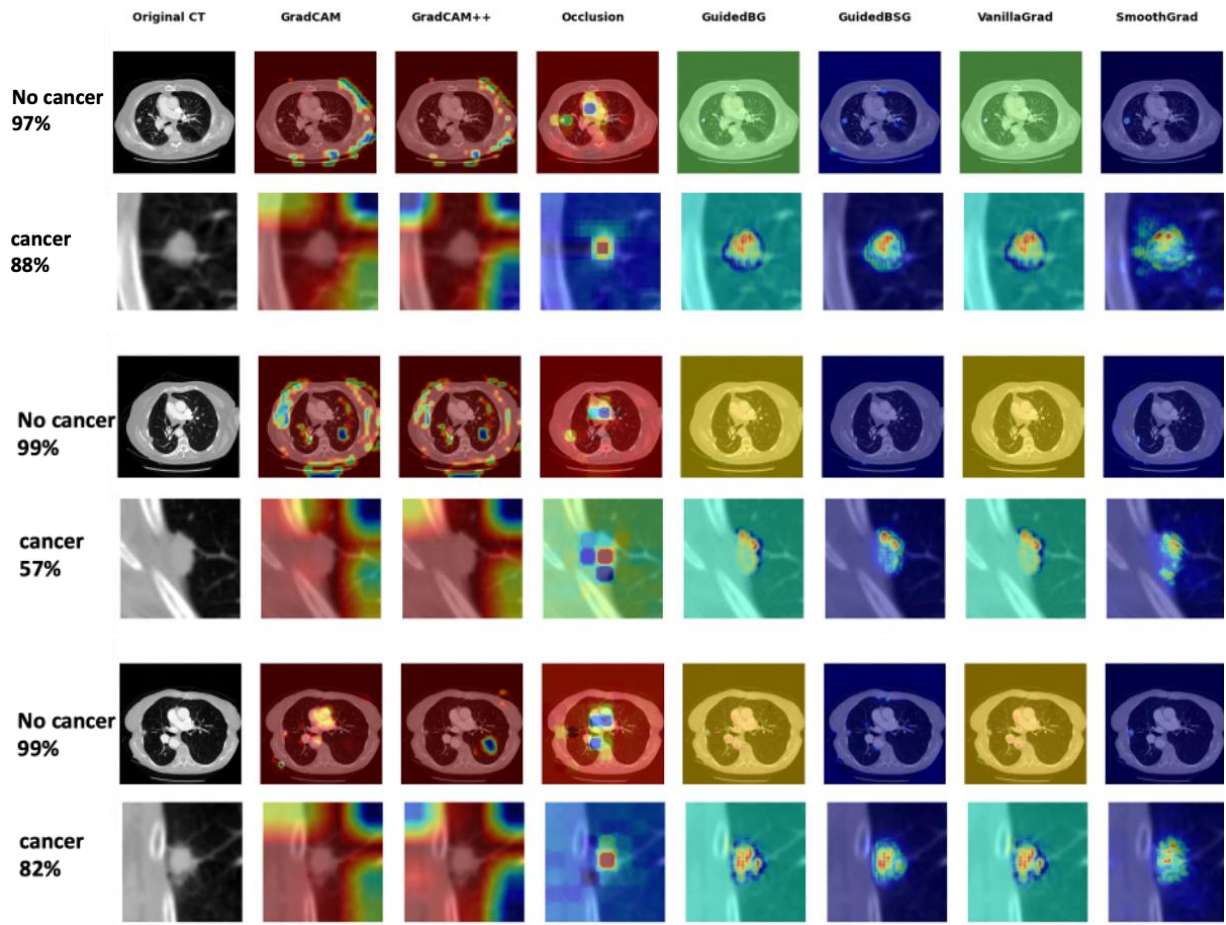

**Figure 6.** Grad-CAM and interpretability visualizations for lesion-level and chest-region models. The figure demonstrates the focus of both models on CT images for malignancy predictions using multiple visualization techniques: Grad-CAM, Grad-CAM++, Occlusion, Guided Backpropagation (GuidedBG), Guided Backpropagation with SmoothGrad (GuidedBSG), VanillaGrad, and SmoothGrad. **Top row:** Chest-region model visualizations, highlighting diffuse activations across the chest area with less specific localization to nodules. **Bottom row:** Lesion-level model visualizations, showing concentrated activations around pulmonary nodules. These results demonstrate the lesion-level model's superior ability to focus on relevant diagnostic features.